\documentclass[12pt]{article}

\usepackage{epsfig}
\usepackage{dcolumn}
\usepackage{amstex}
\usepackage{cite}

\makeatletter \@addtoreset{equation}{section}
\makeatother 
\newcommand{\psibar} {\bar{\psi}}
\newcommand{\scr}[1] {\mbox{\scriptsize #1}}
\newcommand{\mq} {\ensuremath{m_{\scr{q}}}}
\newcommand{\DT} {\ensuremath{\Delta(T)\times T^{4}}}

\newcommand{\tc} {\ensuremath{T_{\scr{c}}}}

\begin{document}
\thispagestyle{empty}
%
 \mbox{} \hspace{1.0cm}
        \today 
 \mbox{} \hfill BI-TP 96/28\hspace{1.0cm}\\
 \mbox{} \hfill IFUP-TH-51/96\hspace{1.0cm}\\
 \mbox{} \hfill hep-ph/9608482\hspace{1.0cm}\\

\begin{center}
\vspace*{1.0cm}
{{\Large\bf   The Temperature Dependence of the $SU(N_{\scr{c}})$ Gluon 
Condensate from Lattice Gauge Theory 
\\}}
\vspace*{1.0cm}
{\large Graham Boyd$^{1,2}$ and David E. Miller$^{1,3}$
\\}
\vspace*{1.0cm}
${}^1$ {Fakult\"at f\"ur Physik, Universit\"at Bielefeld, Postfach 100131, \\
D-33501 Bielefeld, Germany}\\
${}^2$ {Dipartmento di Fisica, Universita di Pisa, Piazza Torricelli 2, \\
I-56100 Pisa, Italy (present address)}\\
${}^3$ {Department of Physics, Pennsylvania State University,
Hazleton Campus, \\
Hazleton, Pennsylvania 18201, USA (permanent address)\\}
\end{center}
\vspace*{2cm} {\large \bf Abstract \\} An analysis of the temperature
dependence of the leading contributions to the gluon condensate for
$SU(N_{\scr{c}})$ lattice gauge theory is presented using the data from recent
Monte Carlo simulations.  The gluon condensate is calculated directly from the
new lattice calculations of the interaction measure.  It is shown how these
computations provide a simple picture for the melting of the condensate around
the deconfinement temperature, and the fact that it is negative, and increases
in magnitude, above \tc. We close with a discussion of the implications for
full QCD of recent results from simulations including fermions.
\newpage


\section{Introduction}

We discuss here the consequences for the gluon condensate at 
finite temperature of the recent high precision lattice results
for the equation of state in $SU(2)$~\cite{Eng4} and
$SU(3)~$\cite{Boyd}, and compare with various other calculations of the
expected high temperature behavior of the condensate.
The relationship between the gluon condensate and equation of state arises due to the scale variance of
quantum chromodynamics (QCD), the trace anomaly, which relates
the trace of the energy momentum tensor
to the square of the gluon field strenth
through the renormalization group beta function.
We expand on the work presented in~\cite{Mill}, 
where the consequences of the new finite temperature 
lattice data for $SU(N_{\scr{c}})$ gauge theory for the gluon condensate were discussed.

These ideas have been 
studied for finite temperature by Leutwyler~\cite{Leut} in
relation to the problems of deconfinement and chiral symmetry.
He discusses in detail the relationship between the
trace anomaly and the gluon condensate, based on the interaction between
Goldstone bosons in chiral perturbation theory.
The condensate has also been investigated directly on the lattice using
plaquette operators by~\cite{Lee,Cam3}.

The energy momentum tensor at finite temperature $T^{\mu\nu}(T)$ can be separated into the 
zero temperature part, $T^{\mu\nu}_{0}$, and the finite temperature
contribution $\theta^{\mu\nu}(T)$:
\begin{equation}
  \label{eq:emtensor}
  T^{\mu\nu}(T) = T^{\mu\nu}_{0} + \theta^{\mu\nu}(T) .
\end{equation}
The zero temperature part, $T^{\mu\nu}_{0}$, has the standard problems with
infinities of any ground state, and so is not readily calculable on the
lattice. The finite temperature part, which is zero at zero temperature, is free of such
problems, and the diagonal elements of $\theta^{\mu\nu}(T)$ are calculated in a
straightforward way on the lattice. The trace  $\theta^{\mu}_{\mu}(T)$ is
connected to 
the thermodynamic contribution to the energy density $\epsilon$ and pressure
$P$  for 
relativistic fields and relativistic hydrodynamics~\cite{LaLi}:
\begin{equation}
  \theta^{\mu}_{\mu}(T) = \epsilon - 3P .
  \label{eq:eps-ideal}
\end{equation}

There are no other contributions to the trace for QCD on the lattice. 
The heat conductivity is zero,
as there is no non-zero conserved quantum number, and there is no velocity
gradient in the lattice study, hence no contribution from viscosity terms.

The dimensionless interaction measure \DT is equal to
the thermal ensemble expectation value of $(\epsilon - 3P)/T^4$. So from 
equation~\eqref{eq:eps-ideal} 
above the interaction measure and the expectation value of the temperature
dependent part of the energy momentum tensor are linked by
\begin{equation}
  \DT = \theta^{\mu}_{\mu}(T).
\label{eq:trace}
\end{equation}

An $SU(N_{\scr{c}})$ calculation of \DT on a lattice of size
$N_{\sigma}^{3}\times N_{\tau}$ with lattice spacing $a$ proceeds
as follows (see~\cite{Eng4,Boyd} for further details). From the action
expectation value at zero temperature, $P_{0}$, as well as the spatial and
temporal action expectation values at finite temperature, $P_{\sigma}$ and
$P_{\tau}$ respectively and $N_{\tau}$ the number of temporal steps, the 
dimensionless interaction measure $\Delta(T)$ \cite{Eng3} is given by:
\begin{equation}
  \Delta(T) = -6N_{\scr{c}}{N^{4}_{\tau}}a\frac{dg^{-2}}{da}\left[
                                2P_{0} - (P_{\sigma}+P_{\tau})
                                \right].
\end{equation}

The crucial part of these recent calculations is the use of the full lattice beta
function, $\beta_{\scr{fn}}=adg^{-2}/da$ to obtain the scale $a$ 
from the bare coupling $g^{2}$~\cite{Eng2,Eng3}. The physical value of $a$ is
then fixed via the string tension.
Without this accurate information on the
temperature scale in lattice units it would not be possible
to make any claims about the behavior of the gluon condensate\footnote{The
  temperature scale in physical units will have an error due to the error on
  the experimental value of the string tension used to connect the lattice
  units to physical units.}. 

Let
$G^{\mu\nu}_a$, where $a$ is the color index for $SU(N_{\scr{c}})$,
denote the gluon field strength tensor.
The quantity $G^2$ is then defined~\cite{Leut}, using the 
beta function ${\beta_{\scr{fn}}(g)}$ with $g$ the bare coupling, to be
\begin{equation}
G^{2}={{-\beta(g)}\over{2g^3}} G^{{\mu}{\nu}}_{a}
            G_{{\mu}{\nu}}^{a}. 
\end{equation}

For a scale invariant system, such as a 
gas of free massless particles,
the trace of the energy momentum tensor, equation~\eqref{eq:trace}, is zero.
A system that is scale variant, perhaps from a particle mass, 
has a finite trace, with the value of the 
trace measuring the magnitude of scale breaking.
At zero temperature it 
has been well understood from Shifman et al.~\cite{SVZ1}
how in the QCD vacuum the trace of the energy momentum tensor
relates to the gluon field strength squared,
$\langle G^2 \rangle_0 $.
A finite temperature gluon
condensate
$\langle G^2 \rangle_{T}$, 
related to the degree of scale breaking at all temperatures, can be
defined to be equal to the trace. 
As the scale breaking in QCD occurs explicitly at all orders in a loop
expansion, the thermal average of the 
trace of the energy momentum tensor, and hence the gluon condensate, need not go
to zero above the deconfinement transition. 
Using the temperature dependent part 
of the trace, and the value of the condensate at zero temperature,
the finite temperature gluon condensate can be written 
\begin{equation}
\langle G^2 \rangle_{T} = 
\langle G^2 \rangle_0 - {\theta}^{\mu}_{\mu}(T),          
\label{eq:condef}
\end{equation}
where the brackets with the subscript $T$ means thermal average.

In the next two sections
we evaluate the gluon condensate as defined in equation~\eqref{eq:condef} above
using the lattice data on ${\theta}^{\mu}_{\mu}(T)$ in 
$SU(2)$ and $SU(3)$ gauge theories. This is followed by
simple melting model for the gluon condensate
in pure gauge theories. After this we look into the high
temperature behavior of $\Delta(T)$ in order to determine the
reliability of its thermodynamical properties in relation to some
earlier models. 
We close the body of the paper with a discussion of the gluon condensate in
full QCD, where the fermions may change the behavior.
Finally we conclude with an evaluation of the present
situation regarding the properties of condensates in the presence
of strong interactions.



\section{ $SU(2)$ pure gauge theory }

In this section we present our evaluations for the gluon
condensate for the $SU(2)$ lattice gauge theory.
Taking the published data \cite{Eng4} for $\Delta(T)$, and using
equations~\eqref{eq:condef} and~\eqref{eq:trace}
we obtain the gluon condensate $\langle G^2 \rangle_T$. Since the 
lattice is not yet able to unambiguously obtain  the condensate at
zero temperature, but only calculates the contribution due
to finite temperature, we need an additional input for $\langle G^2 \rangle_0$ to set
the scale at zero temperature. Although there is some numerical 
uncertainty in the exact value of $\langle G^2 \rangle_0$, it has no
direct influence on the high temperature properties of the gluon
condensate \cite{Leut}.

\begin{table}[t]
\begin{center}
\caption{The lattice results for SU(2) from~\protect\cite{Eng4} and 
  for SU(3) from~\protect\cite{Boyd} for the interaction measure in the
  continuum limit of pure gauge theory. Both the value at the phase transition
  and at the peak are given.}
\label{tab:intmeas}
\begin{tabular*}{0.9\textwidth}{
                c@{\extracolsep{\fill}}
                l
                D{.}{.}{4}
                D{.}{.}{3}
                D{.}{.}{6}
                }
\hline
\multicolumn{1}{c}{} &
\multicolumn{1}{c}{} &
\multicolumn{1}{c}{$T$(GeV)} &
\multicolumn{1}{c}{$\Delta(T)$} &
\multicolumn{1}{c}{\DT} \\
\hline
                                   &$T_{c}$          & 
  0.290 & 0.381 & 0.00269 \\
\raisebox{1.6ex}[0cm][0cm]{$SU(2)$}&$T_{\scr{peak}}$ &
  0.341 & 1.059 & 0.0143 \\
                                   &$T_{c}$          & 
  0.264 & 1.145 & 0.00556 \\
\raisebox{1.6ex}[0cm][0cm]{$SU(3)$}&$T_{\scr{peak}}$ & 
  0.290 & 2.266 & 0.0197 \\
\hline
\end{tabular*}
\end{center}
\end{table}

\begin{figure}[b]
   \begin{minipage}{75mm}
  \begin{center}
    \leavevmode
    \epsfig{file=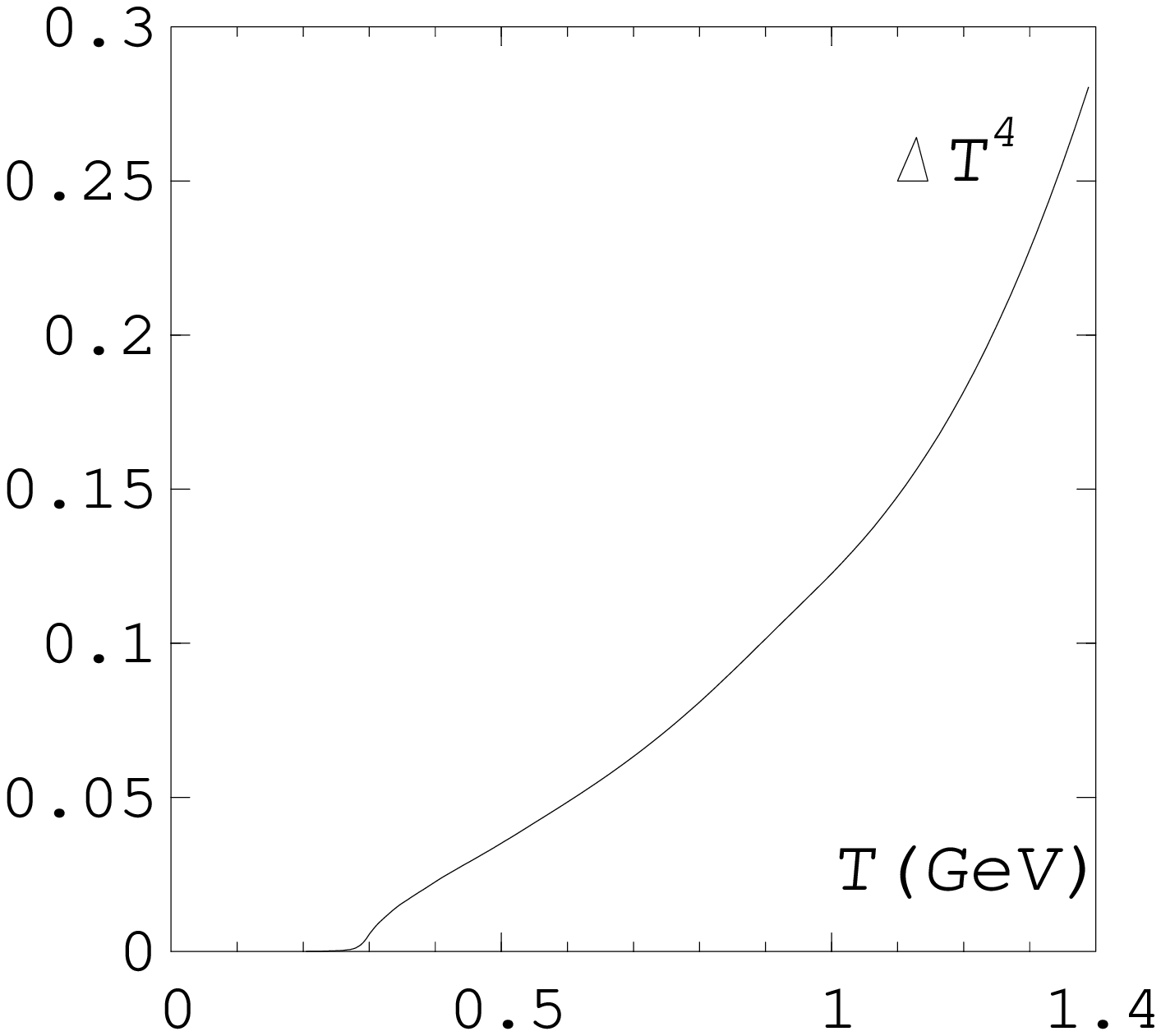
        ,height=85mm
        ,bbllx=85,bblly=200,bburx=550,bbury=650}
    \vskip -79mm
    \epsfig{file=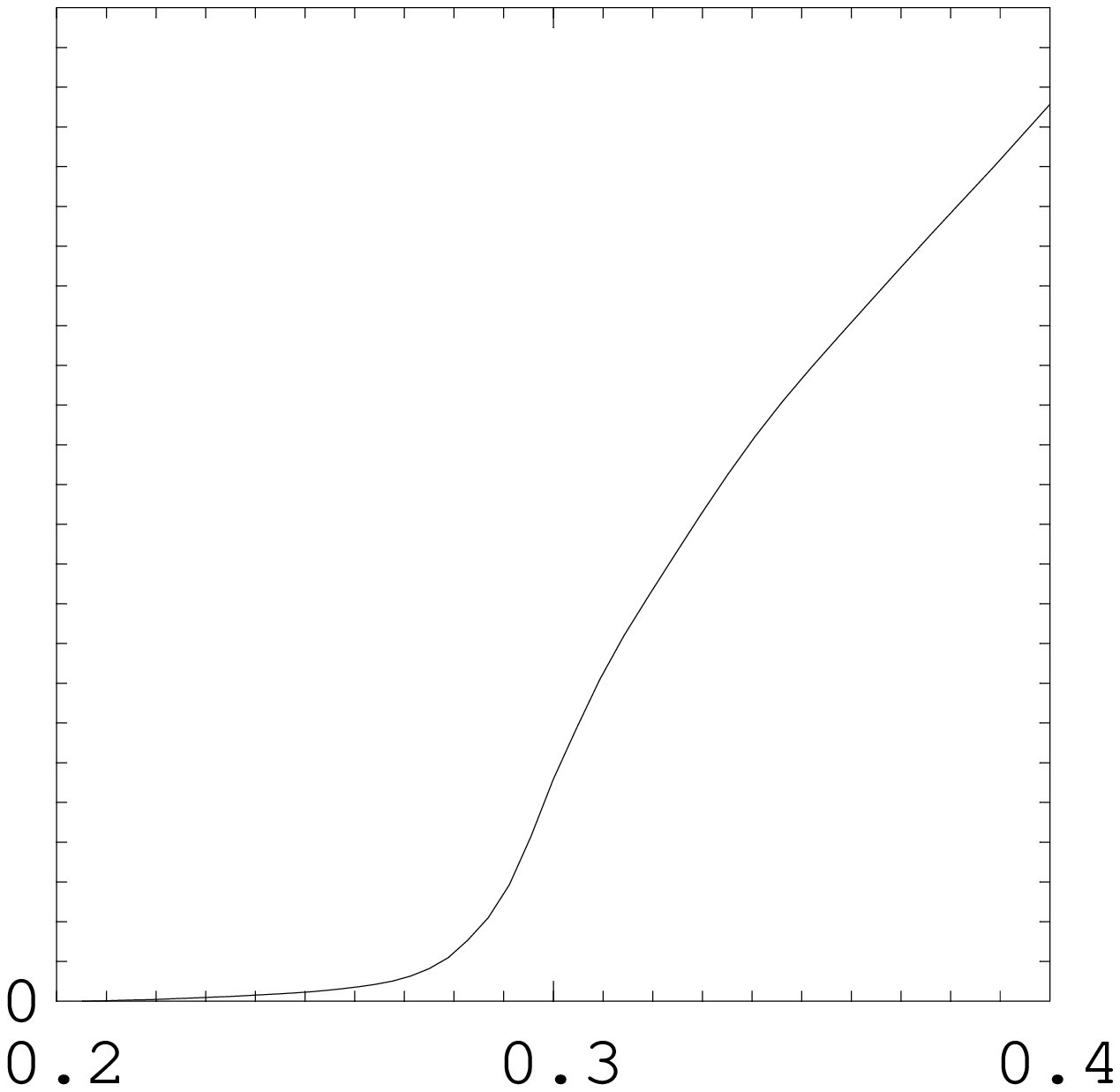
        ,height=48mm
        ,bbllx=160,bblly=200,bburx=580,bbury=650
        }
    \vskip 25mm
   \end{center}
   \end{minipage}
\hfill
\begin{minipage}{75mm}
  \begin{center}
    \leavevmode
      \epsfig{file=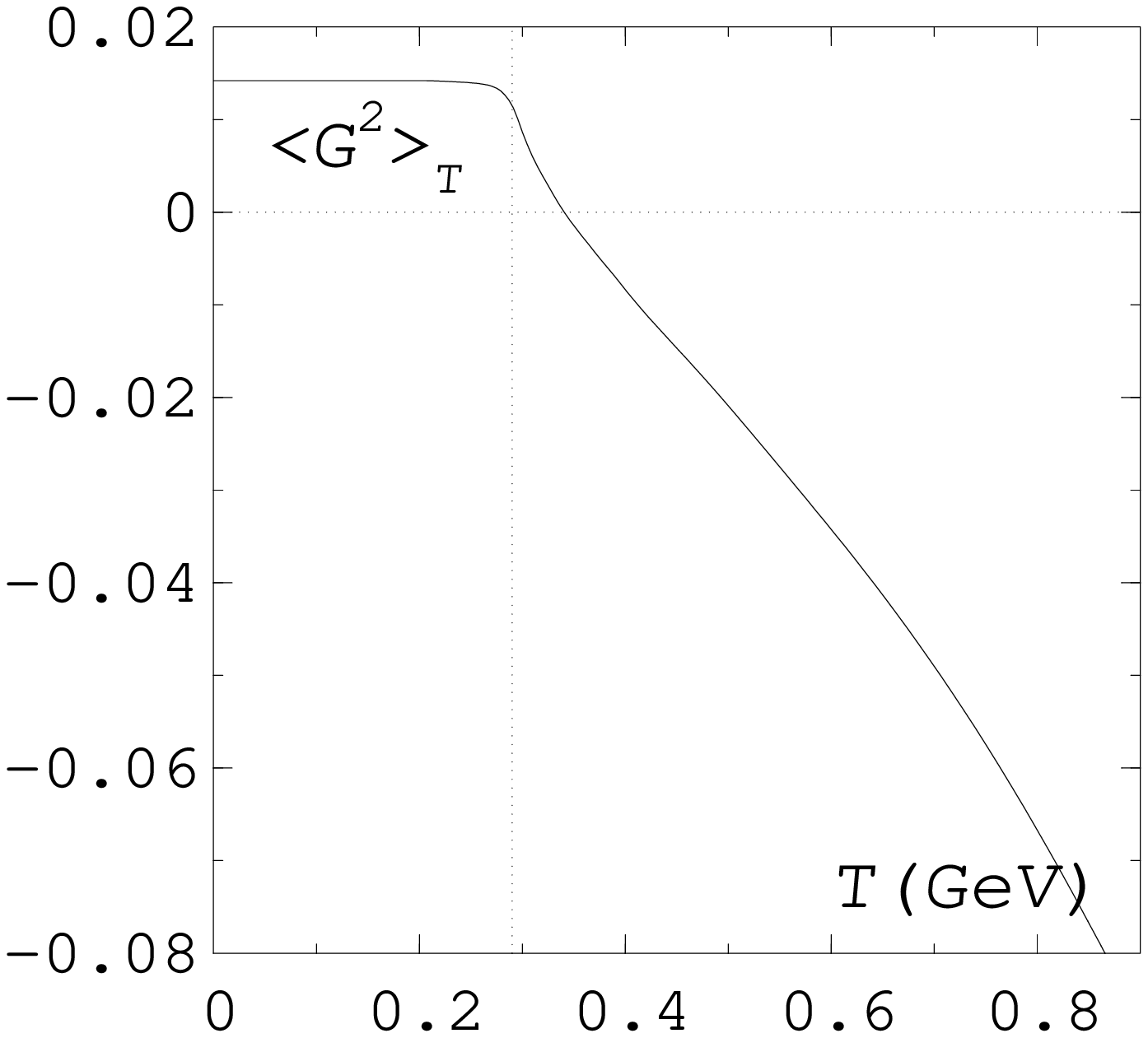,bbllx=85,bblly=200,bburx=550,bbury=650,
          height=85mm}
     \end{center}
   \end{minipage}
   \caption{      \label{fig:cond2}
   Figure~(a) shows \DT\/ for $SU(2)$ gauge theory in terms of $(\text{GeV})^4$.
   The insert shows a magnified region around the phase transition.
   Figure~(b) shows the corresponding condensate. The vertical
   dotted line indicates the critical temperature $\tc = 0.290$~GeV.
          }
\end{figure}

From the lattice data for $SU(2)$ \cite{Eng4} we have
defined the value for $\langle G^2 \rangle_0$
from the peak of $\Delta(T)$ to be
$0.0143\text{GeV}^4$, which is consistent with the range of values
(see the end of section~\ref{sec:meltmodel}
taken from the previous $SU(2)$ computations \cite{Cam1,Cam2,DiG1,DiG2}.
The values corresponding to $\tc$ and $T_{\scr{peak}}$ for $\Delta(T)$ and
$\Delta\times T^4$ are tabulated in table~\ref{tab:intmeas}.                         
Our main reason for this choice for the vacuum value is that the peak of
$\Delta(T)$ represents the point of fastest change of the thermal
properties of the gluon condensate after which the growth of
$\Delta(T)\times T^4$ becomes somewhat slower.  
Using these values we show in Figure~\ref{fig:cond2}(a)
the behavior of $ {\theta}^{\mu}_{\mu}(T)$
over various temperature ranges.
Figure~\ref{fig:cond2}(b) shows the corresponding behavior of the gluon
condensate over a smaller temperature range. We can clearly see that the
gluon condensate goes very rapidly negative for temperatures not
very far above $\tc$. 
note that within the range of numerical values usually considered for
$\langle G^2 \rangle_0$ the temperature at which 
$\langle G^2 \rangle_T$ goes through zero changes little, and remains
above \tc for all these values.
Furthermore, varying $\langle G^2 \rangle_0$ in this range does not cause any
qualitative change in the high temperature behavior of $\langle G^2 \rangle_T$.



\section{ $SU(3)$ pure gauge theory }

Analogous to our approach in the previous section we
use the SU(3) data of~\cite{Boyd} in order to compute \DT. 
After having done this computation we are able to find 
${\theta}^{\mu}_{\mu}(T)$ and thus obtain
$\langle G^2 \rangle_T$ from equation~\eqref{eq:condef}.
Again we have taken the value of \DT\ from the
peak of $\Delta$ to define where $\langle G^2 \rangle_T$
goes to zero. This leads to a value of $0.0197 {(\text{GeV})}^{4}$
for $\langle G^2 \rangle_0$, which is somewhat
larger than the value of $0.012{(\text{GeV})}^{4}$ that is
usually extracted through the sum rules from charmonium decay.
Nevertheless, our value is still within the range of plausible
values~\cite{Dosc}. Furthermore, we should note that for pure gauge theory 
one expects a gluon condensate that is larger than in full QCD.

\begin{figure}[b]
   \begin{minipage}{75mm}
  \begin{center}
    \leavevmode
    \epsfig{file=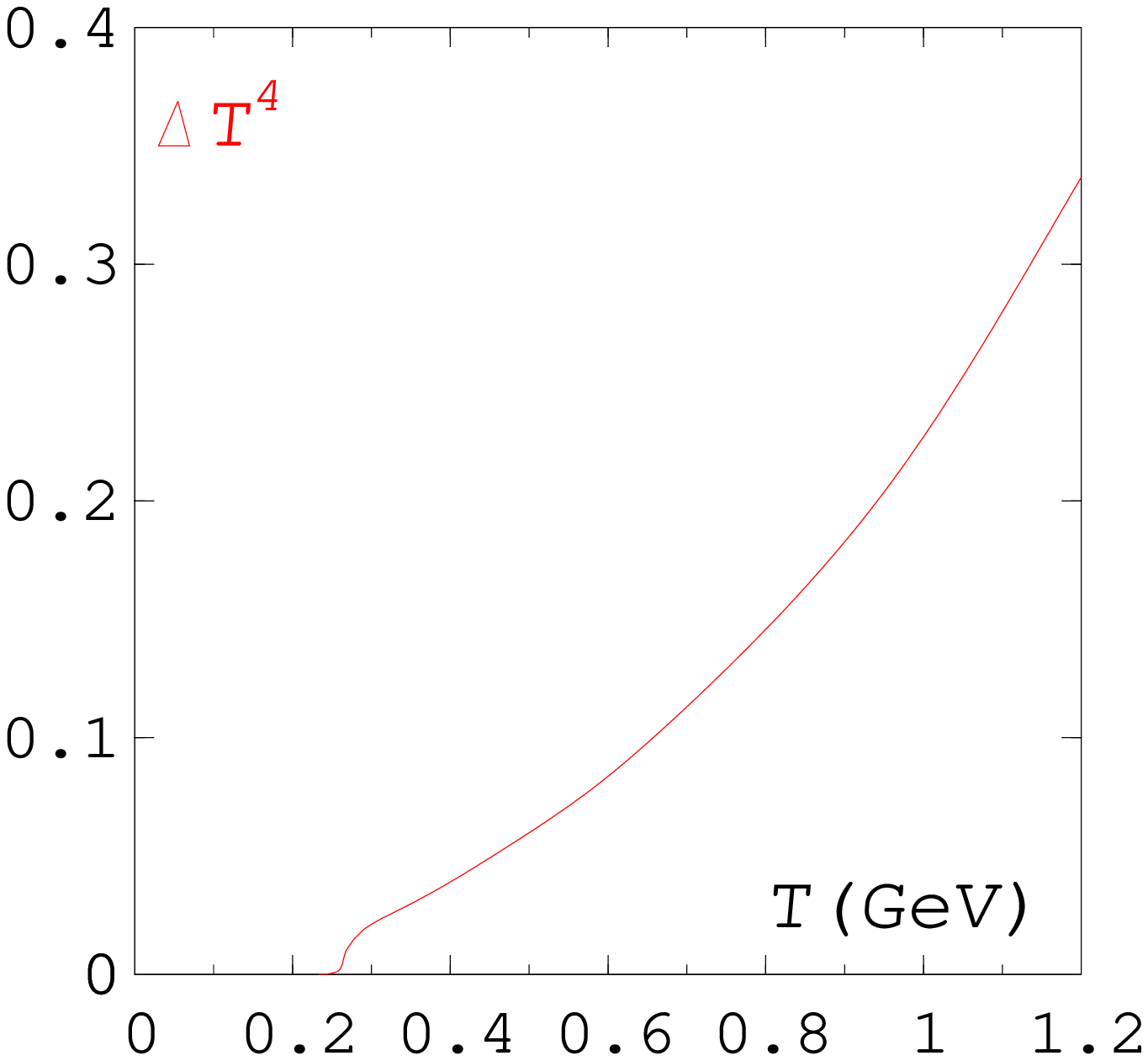
        ,height=85mm
        ,bbllx=85,bblly=200,bburx=550,bbury=650}
   \end{center}
   \end{minipage}
\hfill   
\begin{minipage}{75mm}
  \begin{center}
    \leavevmode
      \epsfig{file=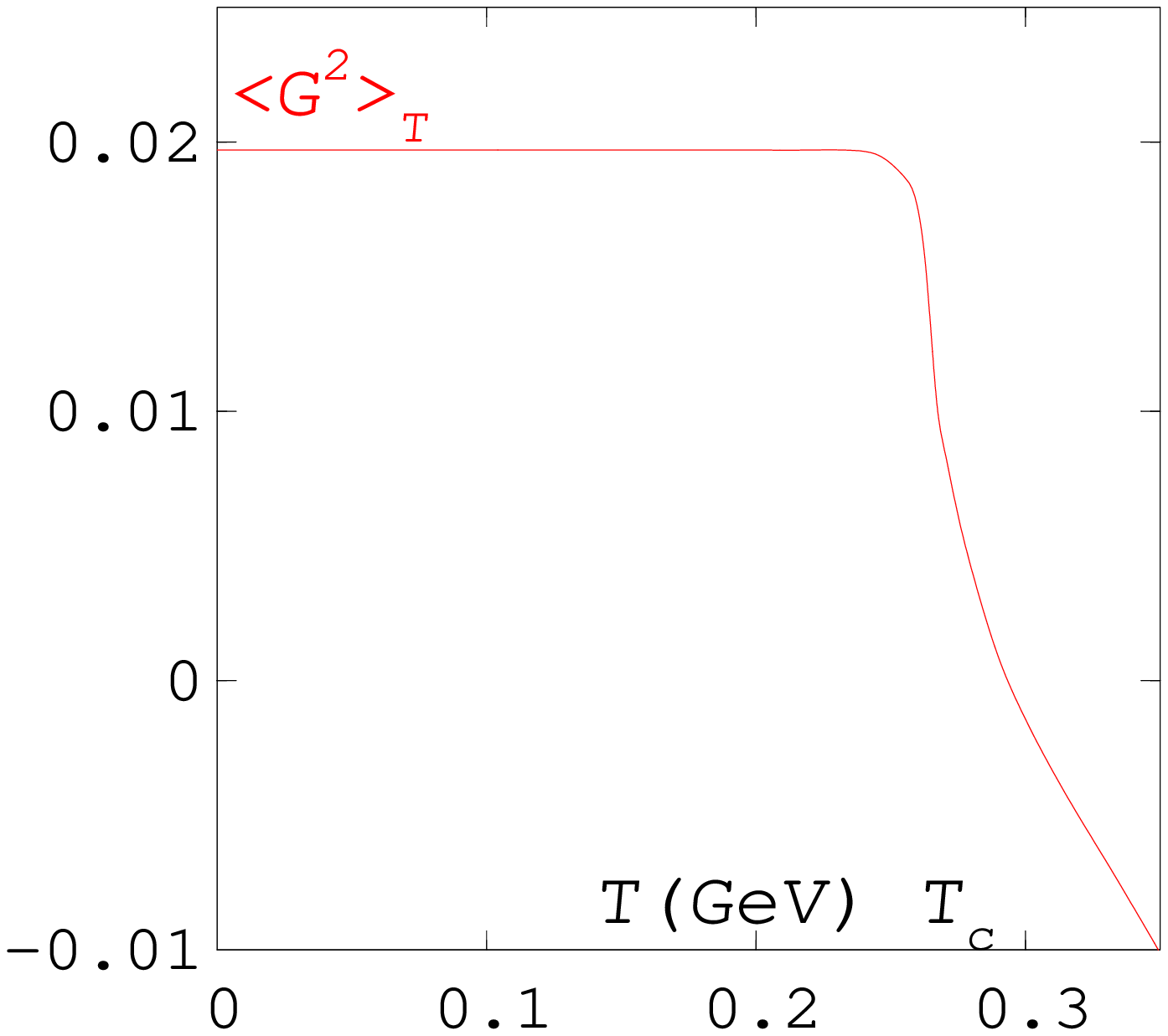,bbllx=85,bblly=200,bburx=550,bbury=650,
          height=85mm}
     \end{center}
   \end{minipage}
   \caption{      \label{fig:cond3}
   Figure~(a) shows $\epsilon - 3P$ for $SU(3)$ gauge theory, extrapolated
   to the continuum limit. 
   Figure~(b) shows the corresponding condensate. The  
   critical temperature is $\tc = 0.264$~GeV.
          }
\end{figure}

The values of \DT\/ are given in table~\ref{tab:intmeas}.
The
plots in Figure~\ref{fig:cond3} show more rapid changes for $SU(3)$ than the
corresponding plot in Figure~\ref{fig:cond2} for $SU(2)$. The qualitative
structure is the same though for both SU(2) and SU(3), both having $\langle G^2
\rangle_T$ dropping rapidly around \tc\/ and a continuation into negative
values at higher temperatures. These results, from a first principles
QCD calculation, are the same as the chiral perturbation theory analysis previously made by
Leutwyler~\cite{Leut}.  
 
The results presented in these two sections differ considerably from 
earlier direct calculations~\cite{Lee,Cam3}. This seems to be due to our 
using a definition that does not require a technically difficult seperation 
of the condensate from a perturbative contribution. Also crucial is the 
use of a fully non-perturbative lattice calculation of the interaction measure,
including the full non-perturbative beta function. 

However, care should be taken in drawing conclusions about the physical
consequences of these results, as the condensate at finite temperature need not
reflect the same physics as at zero temperature.  For example, as 
we will discuss below, there are additional
scale breaking thermal contributions to the condensate which do not
arise from confinement.



\section{Simple Melting Model for the Gluon Condensate}
\label{sec:meltmodel}
 
The results of the previous two sections provide the temperature dependence of
the condensation of gluons for $SU(2)$ and $SU(3)$ lattice gauge theories.
There we defined the condensate to be related to the scale breaking 
at all temperatures, that is the amount by which $\epsilon-3P$ differs from
zero.  The value at $T=0$ is used to set the scale. Other approaches, such as a
direct calculation based on plaquette correlators~\cite{Lee,Cam3,Cam2}, will be
related to this one by renormalisation constants, either additive or
multiplicative, or both.

In Figures~\ref{fig:cond2}(b) and~\ref{fig:cond3}(b)
we have shown $\langle G^2 \rangle_T$ for $SU(2)$  
and $SU(3)$ pure gauge theory respectively. In both cases the general form of this finite
temperature condensate is very much the same. At temperatures well below \tc\/
the vacuum condensate $\langle G^2 \rangle_0$ dominates. From slightly below to
just above \tc\/ there is a sharp drop in the amount of the gluon condensate
present, and a rapid approach to zero. For temperatures above the temperature
where $\langle G^2 \rangle_T$ has reached zero, we see that \DT\/ is
continually growing in absolute value, which causes $\langle G^2 \rangle_T$ to
become increasingly negative.

Thus the processes at lower temperature responsible for `pulling' gluons out of
the condensate continue at higher temperature, driving the condensate further
down to negative values. There are known to be residual interactions in the high
temperature phase, particularly in the spatial (magnetic)
direction~\cite{Boyd,SPAT94}. This leads to a picture of the condensate at
finite temperatures in which the condensate, with zero
total momentum, is composed
of thermal gluon pairs in which one gluon has momentum $p$, the other
$-p$. Since the typical momentum is $p\sim T$, higher temperatures correspond
to both higher momentum states, and a higher density of states, contributing.

Here we shall discuss an oversimplified model with boson fields $B(p)$, and
define the quantity $\langle B^2 \rangle_T$, in analogy to the gluon condensate
at finite temperatures. Here we shall suppose that the oppositely directed
momenta of the two fields are correlated by an interaction in momentum space
$V(|p_1~-~p_2|)$, 
or equivalently, since we have $p_{1}=-p_{2}=p$, $V(2p)$.

Thus the effect of finite
temperature in this model just leads to an expression of the following form:
$\langle B(-p)V(2p)B(+p) \rangle_T$, where the function $V(2p)$ could be any
function, of less than exponential order,
of the magnitude of the momentum. This
could be a positive power of $|p|$, or an algebraic function 
expandable in positive powers of $p$. A modified Bessel function
$K_{\nu}(2p)$ fulfills all the requirements. From this we may
write the equation for the process relating to the vacuum value 
$\langle B^{2}\rangle_{0}$ as
follows:
\begin{equation}
\langle B^2 \rangle_T = \langle B^{2}\rangle_{0} 
+ \langle B(-p)V(2p)B(+p) \rangle_T .
\end{equation}

In order to evaluate the last term in this equation, we expand $V(2p)$ in powers
of the argument $2p$ and approximate the field distributions with a simple
Planck distribution function at finite temperatures $n(+p)$ and $n(-p)$.  The
integral then takes the form
\begin{equation}
\langle B(-p)V(2p)B(+p) \rangle_T = \int {d^{3}p\over{(2\pi)^3}}~n(-p)V(2p)n(+p),
\end{equation}
where the spatial volume has been divided out leaving a density term.
After carrying out the expansion in powers of the momentum indicated above, 
and using the spherical symmetry in momentum space, we may exactly evaluate
this integral to obtain
\begin{equation}
\langle B(-p)V(2p)B(+p) \rangle_T = \frac{-1}{2\pi^{2}} \sum_{j=0}^{\infty}
V_{j}2^{j}T^{3+j}(j+2)!\zeta(j+2),     
\end{equation}
where $\zeta(s)$ is the Riemann zeta function with the argument $s$.        
We now can carry out the numerical evaluation of the above terms, which yields
\begin{equation}
\langle B(-p)V(2p)B(+p) \rangle_T~=~-0.16667V_0 T^3~-0.73076V_1 T^4~-..., 
\end{equation}
where the $V_{j}$ are the coefficients of the expansion of the interaction. As
an example of these results, let us suppose that the only nonzero term in the 
expansion is $V_{0}$, which is just a constant. Then the condensation equation
becomes
\begin{equation}
\langle B^2 \rangle_T=  \langle B^{2}\rangle_{0} - 0.16667V_{0} T^{3} .
\label{eq:sunbec}
\end{equation}
This equation is analogous to the Bose-Einstein condensation (BEC)
equation, with the condensation temperature defined as the cube root of $6
\langle B^{2}\rangle_{0}/V_0$.  However, in contrast to the usual BEC in
statistical particle systems, equation~\eqref{eq:sunbec} applies both above and below the
condensation temperature. Therefore, $\langle B^2 \rangle_T$ can become
negative at sufficiently high temperatures for a system which has only boson
fields present. 
                                     
A reasonable picture of this process seems to be that the 
attractive interactions of the gluons at these very high temperatures
causes a sort of `antiscreening' effect, a continuation of the
melting process to values exceeding that of the vacuum condensate.  Before we
are able to continue this discussion in more detail, we must consider some
points in the analysis of the vacuum contributions to the gluon condensate. The
classical analysis of the gluon contributions at the lowest order in the
operator product expansion \cite{SVZ1} arrives at a value for the $\langle G^2
\rangle_0$ term of about $0.012 \text{GeV}^4$.  Although the numerical
confirmation of this analysis has later shown some difficulties due to certain
singularities, the basic method of the operator product expansion for obtaining
hadron properties from the QCD sum rules is well confirmed \cite{Rein}.  

A very
recent study \cite{Dosc} shows a rather large range of values of $\langle G^2
\rangle_0$ between 0.0127 and 0.0355 $\text{GeV}^4$. Also a number of
computations for the pure lattice gauge theories have been carried out for
$SU(2)$ \cite{DiG1,DiG2,Cam1} as well as for $SU(3)$ \cite{Cam2}. A check of
the range of $G^2$ for $SU(2)$ using the relationship of $\Lambda_L$ to \tc\/
for the new computations on $SU(2)$ \cite{Eng4} with \tc\/ at 0.290
$\text{GeV}$ gives values around 0.0059 to 0.0236 $\text{GeV}^4$. These
values , although somewhat smaller than the QCD values, are still quite
reasonable. The numbers from lattice computations for $SU(3)$ \cite{Cam2} yield
values much too large when the new computations for $SU(3)$ \cite{Boyd}
relating \tc\/ to $\Lambda_L$ are used. In this case we choose the values from
QCD.  The discussion above leading to a melting of the gluon condensate and a
continuation of the same process into the high temperature region brings in the
question of the thermodynamical properties of such a system.  For this reason
we shall look more closely at the high temperature results from lattice gauge
theory.



\section{High Temperature Behavior of $\Delta(T)$}

\begin{figure}[b]
   \begin{minipage}{75mm}
  \begin{center}
    \leavevmode
      \epsfig{file=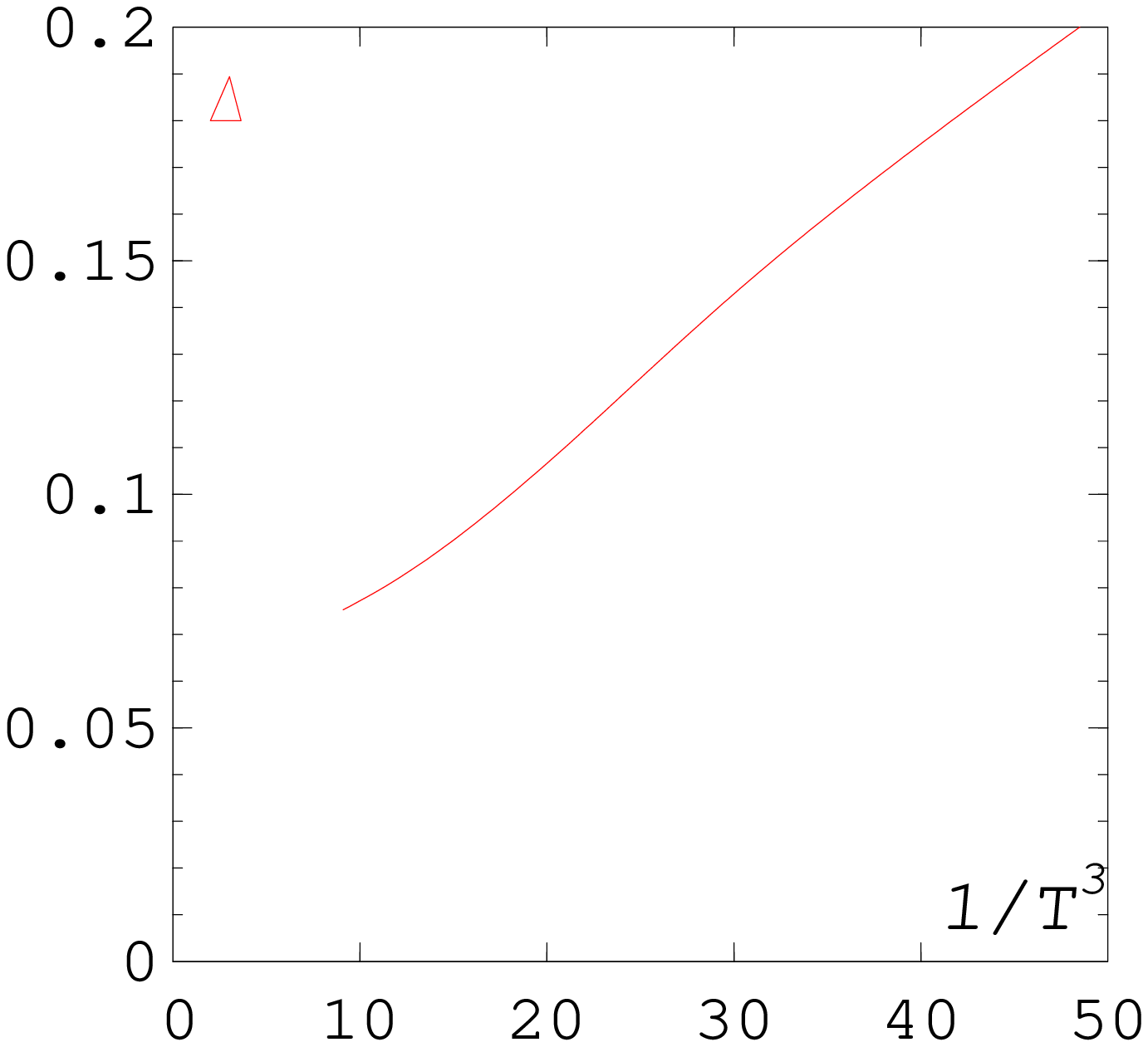,bbllx=85,bblly=200,bburx=550,bbury=650,
          height=85mm}
   \end{center}
   \end{minipage}
\hfill   
\begin{minipage}{75mm}
  \begin{center}
    \leavevmode
      \epsfig{file=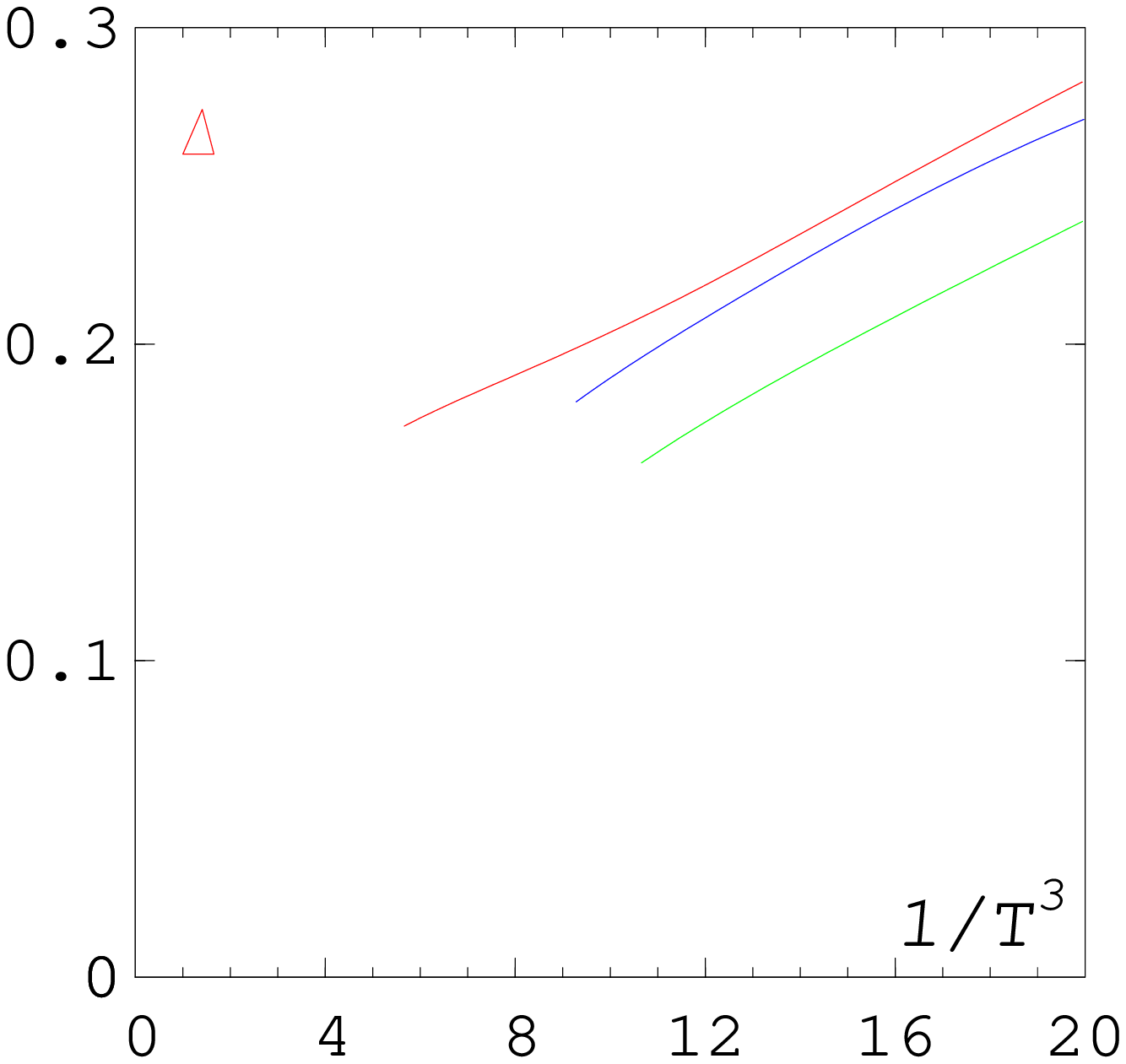,bbllx=85,bblly=200,bburx=550,bbury=650,
          height=85mm}
     \end{center}
   \end{minipage}
   \caption{      \label{fig:del_t3}
Figure~(a) shows $\Delta(T)$ as a function of $1/T^3$ for $SU(2)$ lattice
gauge theory\cite{Eng4},where the scale of the abscissa is ${\tc}^{-3}\times
10^{-3}$;
Figure~(b) shows $\Delta(T)$ for $SU(3)$ with the three lattice sizes \cite{Boyd} 
$16^3\times4$ (top line), $32^3\times6$ (middle line), $32^3\times8$ 
(bottom line). The
scale of the abscissa is the same as for $SU(2)$.   }
\end{figure}

In this section we present a discussion of the properties of $\Delta(T)$ at
temperatures above $2T_c$ up to around $5T_c$. Some of the early work in this
temperature range was done using perturbative estimates by K\"allman
\cite{Kall} and Gorenstein and Mogilevsky \cite{GoMo}. Also Montvey and
Pietarinen \cite{MoPi} looked at the asymptotic properties of the gluon gas.
K\"allman suggested a mean-field type of model to which he fitted the lattice
$SU(2)$ data \cite{Eng1} using a linear temperature dependence of the quantity
\DT\/ above \tc\/. At first appearance the data for $\Delta(T)$ fit rather
well for the temperatures somewhat above the peak. A quite different analysis
was carried out by Gorenstein and Mogilevsky, who compared the behavior of
$\epsilon/T^4$ and $3P/T^4$ as functions of $1/T^3$. Again it appeared that
they approach each other at small values of $1/T^3$ using the best computer data
available at the time \cite{Eng1}. With these new data for $SU(2)$ and $SU(3)$
\cite{Eng4,Boyd} we are able to compare these various approaches to
the high temperature behavior of $\Delta(T)$ at temperatures in the above
range.  Figures~\ref{fig:del_t3}(a) and~\ref{fig:del_t3}(b) show the high temperature behavior of $SU(2)$
and $SU(3)$ respectively, as a function of $1/T^3$.

We see from these plots that the general appearance is much the same in the
different cases. $\Delta(T)$ appears to be a linearly
increasing function of the variable $1/T^3$. In Figure~\ref{fig:del_t3}(b) 
we have provided the errors on the points
so that we may assess the validity of our
fitting procedure on the plots. We can see that for the $16^3\times4$ and the
$32^3\times6$ both the data points and the curves are close with rather small
errors, while for the $32^3\times8$ the highest point is well
away from the interpolated curve, but still consistent within the error.

The question of how well these results can be compared to earlier work may
now be answered. The assumption of glueballs of a mass $m_{gb}$ gives a high
temperature form for $\Delta(T)$ which is written as  follows~\cite{MoPi}:
\begin{equation}
\Delta(T)~=~{d_{gb}\over{6{\pi}^2}}{(m_{gb}/T)^3}K_1(m_{gb}/T),               
\end{equation}
where $d_{gb}$ is the statistical degeneracy of the glueball states.
In the high temperature limit for this case
$\Delta(T)$ is a much more slowly rising function of $1/T^3$ than for the
numerical results shown in the above plots. The asymptotic behavior of the
form of $\Delta(T)$ assumed by K\"allman \cite{Kall} and Gorenstein and
Mogilevsky \cite{GoMo} is qualitatively similar to ours in the region where 
data exists. However, the latter \cite{GoMo} present their data in
terms of energy density and the pressure separately, and indicate that the two curves
converge at small values of $1/T^3$. From our plot
we see that this behavior is not supported by the present
lattice data, rather there exists a considerable gap with no
indication of $\Delta(T)$ approaching zero at small $1/T^3$. 

If one were
to consider a simple bag type of
model with the bag constant $B$ independent of the temperature, one finds
immediately that $\Delta(T)$ is proportional to $1/T^4$.
This dependence clearly results in a too rapid rise in $\Delta(T)$ as a
function of $1/T^3$, which is not consistent with the data shown in Figure~\ref{fig:del_t3}.
plots. Although we are not able to exactly determine the asymptotic form of
the function at high temperature, it appears to scale linearly with $1/T^{3}$
and to be non-zero at high temperature, ie., $\epsilon-3p>0$, for
temperatures between $2T_c$ and $5T_c$.



\section{Extension to full QCD}

Finally we would like to discuss the changes due to the presence of dynamical
quarks with a finite mass.  There have been recent calculations of the
thermodynamical quantities in full QCD with two 
flavours~\cite{BKT94,MILCeos6}, and with four flavours~\cite{edwinqmfklat96} of staggered
quarks.
These calculations are not yet as accurate as those in pure gauge theory for
two reasons. The first is the prohibitive cost of obtaining statistics similar
to those obtained for pure QCD. So the error on the interaction measure is
considerably larger. The second reason, perhaps more serious, lies in the
effect of the quark masses currently simulated. They are relatively heavy,
which increases the contribution of the quark condensate term 
$\mq\langle\psibar\psi\rangle$ to the interaction measure.

The transition
takes place when the temperature is high enough to excite many thermal mesons.
So if the lightest mesons are heavier than in reality, due to the quark mass
being heavier, the transition
temperature should be higher.  However, the lattice spacing is obtained by
imposing the experimental rho mass, $m_{\rho}=770$MeV, on the $\rho$ for every
quark mass simulated.
This means that the
temperature scale in physical units is itself not correct, and may well change considerably near
$\tc$ when smaller quark masses are simulated.
The condensates in full QCD have also been considered by Koch and
Brown~\cite{kochbrown}, but the lattice measurements used were not
fully non-perturbative, nor was the temperature scale
obtained from the full non-perturbative beta-function. 

\begin{figure}[tb]
  \begin{center}
    \leavevmode
    \epsfig{file=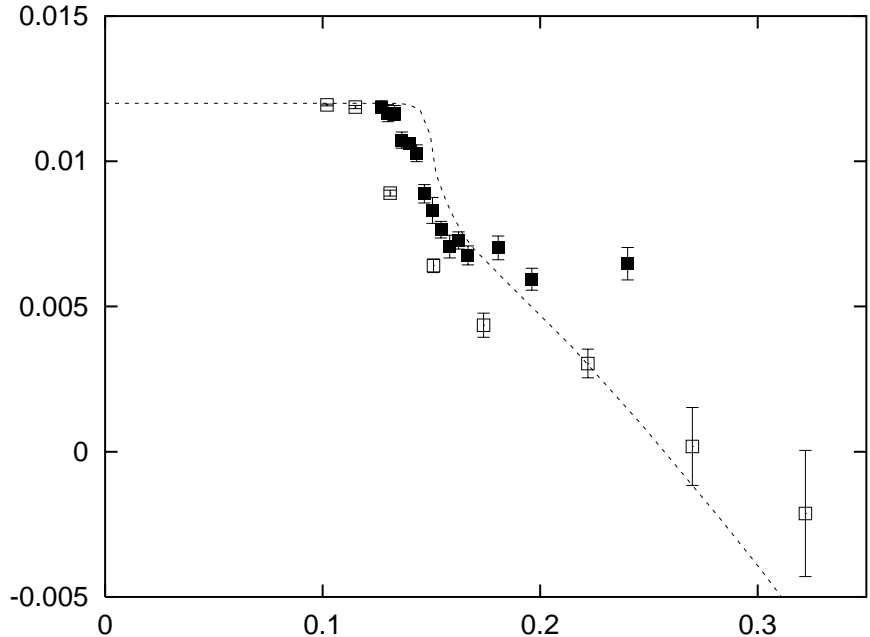,height=90mm}
    \caption{The gluon condensate in full QCD. The solid squares come from the
    two flavor~\protect\cite{MILCeos6}, the open squares are for four
    flavous~\protect\cite{edwinqmfklat96}. The dashed line shows the
    pure gauge condensate
    rescaled by the number of degrees of freedom.
    The units are the same as the previous figures.}
    \label{fig:full}
  \end{center}
\end{figure}

In Figure~\ref{fig:full} we have plotted the condensate from the $N_{\tau}=6$
data of~\cite{MILCeos6} for two flavors with $\mq/T=0.075$, and preliminary
results from Bielefeld~\cite{edwinqmfklat96} for four flavors with $\mq/T=0.2$.
The condensate is obtained from the interaction measure in a type of chiral
limit; the simulation is performed with quarks of the stated mass, but the
energy density is measured with the quark mass set to zero.

We have also indicated with a dashed line the pure gauge theory results of
Figure~\ref{fig:cond3} using
$\tc=0.150\text{GeV}$ to set the scale. The four flavor temperature scale has
been converted to physical units using $\tc=0.130$GeV~\cite{MTC92}.

From this figure it is clear that no conclusion can yet be drawn for full QCD.
The results of the MILC collaboration indicate that the condensate may flatten
off above \tc. This behavior cannot be achieved by simpy rescaling the pure
$SU(3)$ results, and suggests a qualitative change.  If one assumes that the
condensate should go to zero at high temperature, then this indicates that the
zero temperature gluon condensate is not 0.012 GeV$^4$ but rather 0.007 GeV$^4$.
Equivalently, one could claim that the condensate drops to half the zero
temperature value in the high temperature phase.

Unfortunately the MILC data do not go much above \tc, so this conclusion
depends very strongly on the last few points.  The preliminary
Bielefeld data, which go to higher temperatures, behave similarly to the
rescaled pure $SU(3)$ data, and indicate that the condensate does become 
negative at high temperatures.

As mentioned above there are severe difficulties to be overcome when the
temperature scale, coming from the full non-perturbative beta function, is
fixed for full QCD simulations. It is quite plausible that the differences
between the two simulations come from the different temperature scales used, and
that more precise data will remove this difference. 
Also, the Bielefeld group uses an improved action for both the gluon and
fermion parts of the action, while the MILC group uses the standard action for
both. 
So it is clearly too early 
to come to any conclusion about the high temperature behavior of the 
gluon condensate in full QCD. 

Having said that, let us end by speculating on why it may be possible that full  
QCD behave differently from the pure gauge theory. A flattening off of the 
condensate suggests that the system is more stable because of the quarks.
We consider the simple model presented in section~\ref{sec:meltmodel}, but now
with the quarks included. The thermal quarks will change the interactions in
the system due to the screening of the static interactions. Although there are 
still residual interactions in the magnetic sector even though the string tension
itself goes to zero while the spatial string tension remains nonzero, it seems 
reasonable that the momentum dependent interactions that led to a negative 
condensate for pure gauge theory are screened at large distances in full 
QCD so that possibly no negative condensate forms
at least within the range of the data.



\section{Conclusions}
The main conclusions for the temperature dependence of the gluon condensate 
come from the simulations of the $SU(N_{\scr{c}})$ lattice gauge theories at 
finite temperature~\cite{Eng4,Boyd}. These simulations provide not only an
accurate computation of the interaction measure $\Delta(T)$, but also
of the temperature scale due to the calculation of the 
beta function at finite temperature. Both are vital in the
computation of $\theta^{\mu}_{\mu}(T)$ in the previous sections. We have also
given a discussion of these quantities in relation to full QCD.

It is clear that the condensate becomes negative in pure gauge
theory, and keeps dropping with increasing temperature. The point
at which it becomes zero has no special significance. This can be understood in
terms of a model based on boson fields as coming from interactions between the
thermal bosons. 

For full QCD it is not yet possible to draw a conclusion.  There are data
supporting both the condensate going to zero and its becoming negative at high
temperatures. Future simulations will show whether or not pure and full QCD
have qualitatively different behaviours.

Although we are less able to say what happens to the temperature scale
and the beta function in full QCD, we can see from the present
newer computations as well as the known results \cite{BKT94}
that $\theta^{\mu}_{\mu}(T)$ is always positive at all computed values
of the temperature. Therefore, provided that the vacuum contribution to
the trace of the energy momentum tensor is positive, we may conclude from
Equation (1.1) that the total trace of $T^{\mu}_{\mu}(T)$ always remains
positive. In this case the divergence of the dilatation and conformal currents
remains positive. Thus these currents are not conserved. Therefore, the scale 
symmetry remains broken at all temperatures for nonabelian lattice gauge theories. 



\section{Acknowledgements}

\medskip The authors would like to thank Jochen Fingberg, Frithjof Karsch,
Krzysztof Redlich and Gennady Zinovjev for helpful discussions.
We are especially grateful to the Bielefeld group and the MILC collaboration
for providing us with their data, and to J\"urgen Engels for the use of his
programs and many valuable explanations of the lattice results. One of us (DEM)
would like to express his appreciation to the Fakult\"at f\"ur Physik der
Universit\"at Bielefeld for the hospitality in the friendly and creative
atmosphere.  This work was partly funded (for GB) by the European Union {\em
  Human Capital and Mobility} program HCM-Fellowship contract ERBCHBGCT940665.



\end{document}